\documentclass[aps,prb,  twocolumn,showpacs,preprintnumbers,superscriptaddress]{revtex4}

\usepackage{graphicx}
\usepackage{dcolumn}
\usepackage{bm}

\begin{document}

\title{Anomalous pressure effect on the remanent lattice striction of a (La,Pr)$_{1.2}$Sr$_{1.8}$Mn$_{2}$O$_{7}$ bilayered manganite single crystal}

\author{M.Matsukawa} 
\email{matsukawa@iwate-u.ac.jp }
\author{A.Tamura}
\affiliation{Department of Materials Science and Technology, Iwate University , Morioka 020-8551 , Japan }
\author{S.Nimori}
\affiliation{National Institute for Materials Science, Tsukuba 305-0047 ,Japan}
\author{R.Suryanarayanan}
\affiliation{Laboratoire de Physico-Chimie de L'Etat Solide,CNRS,UMR8648
 Universite Paris-Sud, 91405 Orsay,France}
\author{T.Kumagai}
\author{Y.Nakanishi}
\affiliation{Department of Materials Science and Technology, Iwate University , Morioka 020-8551 , Japan }
\author{M.Apostu}
\author{A.Revcolevschi}
\affiliation{Laboratoire de Physico-Chimie de L'Etat Solide,CNRS,UMR8648
 Universite Paris-Sud, 91405 Orsay,France}
\author{K. Koyama}
\author{ N. Kobayashi}
\affiliation{Institute for Materials Research, Tohoku University, Sendai  
980-8577, Japan}

\date{\today}

\begin{abstract}
We have studied the pressure effect on magnetostriction, both in the $ab$-plane and along the $c$-axis , of a  (La,Pr)$_{1.2}$Sr$_{1.8}$Mn$_{2}$O$_{7}$ bilayered manganite single crystal over the temperature region where the field-induced ferromagnetic metal (FMM) transition takes place. 
For comparison, we have also examined  the pressure dependence of  magnetization curves  at the corresponding temperatures.
The applied pressure reduces  the critical field of the FMM transition  and it  enhances the remanent magnetostriction. 
An anomalous pressure effect on the remanent lattice relaxation is observed and is similar to the pressure effect on the remanent magnetization along the $c$-axis. 
These findings are understood from the view point that 
the  double-exchange interaction driven  FMM state is strengthened by application of pressure.

\end{abstract}

\renewcommand{\figurename}{Fig.}
\maketitle
\section{ INTRODUCTION}
The observation of a phenomenon of a colossal magnetoresistance (CMR) effect has renewed the interest for doped manganites  with perovskite 
structure \cite{TO00}. Though the insulator to metal (IM) transition and its associated CMR are well explained  
on the basis of  the double exchange (DE) model,  it was pointed out that the dynamic Jahn-Teller 
(JT) effect due to the strong electron-phonon interaction, plays a significant role in 
the appearance of CMR as well as of the DE interaction \cite{ZE51,MI95}.  
Furthermore, Dagotto 
et al proposed a phase separation model where  ferromagnetic (FM) 
metallic and antiferromagnetic (AFM) insulating clusters coexist, which strongly
 supports recent experimental studies of the physics of manganites  \cite{DA01}. 
 
Moritomo et al. have reported that the La$_{1.2}$Sr$_{1.8}$Mn$_{2}$O$_{7}$ bilayered manganite  exhibits a paramagnetic insulator (PMI) to ferromagnetic metal (FMM) transition around $T_{c}\sim$120K  and an associated CMR effect \cite{MO96}. The Pr(Nd)-substitution on the La-site leading to $\{$La$_{1-z}$,Pr(Nd)$_{z}\}$ $_{1.2}$Sr$_{1.8}$Mn$_{2}$O$_{7}$  causes an expansion along the $c$-axis but a shrinkage along the a(b) axis, resulting in a change of the $e_{g}$ electron occupation from the $d_{x^2-y^2}$ to the  $d_{3z^2-r^2}$ orbital \cite{MO97,OG00}. These findings accompany not only  a suppression of the PMI to FMM transition temperature, $T_{c}$,  but also a variation of the easy axis of magnetization from the $ab$-plane to the $c$-axis. For the $z=0.6$ crystal,  the spontaneous ferromagnetic metal phase disappears at the ground state but the field-induced PMI to FMM phase is obtained over a wide range of temperatures. The magnetic phase diagram in the ($H,T$) plane, established from  magnetic measurements carried out on the $z=0.6$ crystal, is presented in Fig.1,  with three regions labeled as PMI, FMM , and mixed states (white area). The white area is characterized by a hysteresis in the magnetization curves. 
Application of physical pressure is a powerful tool to investigate the lattice effect on magnetic and electronic properties of doped manganites as well as the \textit{chemical pressure} effect due to the other rare-earth ion substitution on the La site\cite{HW95}.  
Many studies have been carried out so far on the  effect of pressures on structural, magnetic and transport properties in bilayered manganites \cite{AR97L,KI97,ZH98}. 
In half doped bilayered manganite La$_{2-2x}$Sr$_{1+2x}$Mn$_{2}$O$_{7}$($x$=0.5), near the $x$=0.4 crystal studied here,
the long-range orbital- and charge-ordered state appears over a limited temperature range between 100 and 210 K \cite{AR00}. 
This finding is taken to be related to the polaronic state of the optimally doped crystal exhibiting the CMR effect through an orbital frustration
 in the PI phase \cite{AR02}.  
We report here the  effect of pressure on magnetostriction, both in the $ab$-plane and along the $c$-axis, associated with the field-induced 
IM transition of a  (La$_{1-z}$,Pr$_{z}$)$_{1.2}$Sr$_{1.8}$Mn$_{2}$O$_{7}$ bilayered manganite single crystal.  
Just after removing the field, the system still remains in a metastable FMM state and it then comes back to the original PMI state through a mixed state made of FMM and PMI regions. 
Here, the FMM state lies in a local energy minimum and the PMI state is located in a global minimum of free energy, when 
the applied field is switched off.
Next, we examine the giant pressure effect observed  on the relaxation time of  remanent magnetostriction. Furthermore, we compare the present results 
with the magnetic relaxation data on the $z=0.6$ crystal. 

\begin{figure}[ht]
\includegraphics[width=9cm,clip]{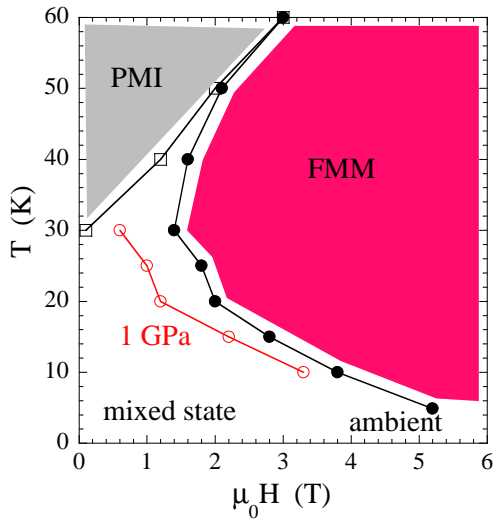}%
\caption{(Color online)Magnetic phase diagram in the ($H,T$) plane established from the magnetic measurements carried out 
on the $z=0.6$ crystal ($H\parallel c$).  Three regions are distinguished and labeled as PMI, FMM, and mixed states (white area). 
The white area is characterized by a hysteresis in the magnetization curves. The phase transition lines between PMI( or mixed phase) 
and FMM are defined as a maximum of $dM/dH$. For comparison, we show the PMI to FMM phase transition line under a pressure of 1 GPa  
determined from the pressure data of $MH$ curves, as shown in the text. }
\label{Phase}
\end{figure}%

\section{EXPERIMENT}
Single crystals of (La$_{0.4}$,Pr$_{0.6}$)$_{1.2}$Sr$_{1.8}$Mn$_{2}$O$_{7}$ were grown by the floating zone method using  a mirror furnace. 
The calculated lattice parameters of the tetragonal crystal structure of the crystals used here were shown in a previous report\cite{AP01}. 
 The dimensions of  the $z$=0.6 sample are 3.4$\times$3 mm$^2$ in the $ab$-plane and 1mm along the $c$-axis.  Measurements of magnetostriction , 
both in the $ab$-plane and along the $c$-axis , were done by means of a conventional strain gauge method at the Tsukuba Magnet Laboratory, 
the National Institute for Materials Science (NIMS) and at the High Field Laboratory for Superconducting Materials, Institute for Materials Research, 
Tohoku University. First, the sample was cooled down to the selected temperatures in zero field and  we then started  measuring the isothermal
magnetostriction upon increasing (or decreasing) the applied fields at a sweep rate of  0.2 T/min. 
Finally, we recorded the isothermal remanent magnetostriction as a function of time just after switching off the field.
Hydrostatic pressures  in both magnetostriction and magnetization experiments were applied  by a clamp-type cell using Fluorinert as a pressure transmitting medium. The pressure was calibrated by the critical temperature of lead. 
The magnetization measurements were made  with a superconducting quantum interference device  magnetometer both at Iwate University and  NIMS.

\section{Results and discussion}
\begin{figure}[ht]
\includegraphics[width=9cm]{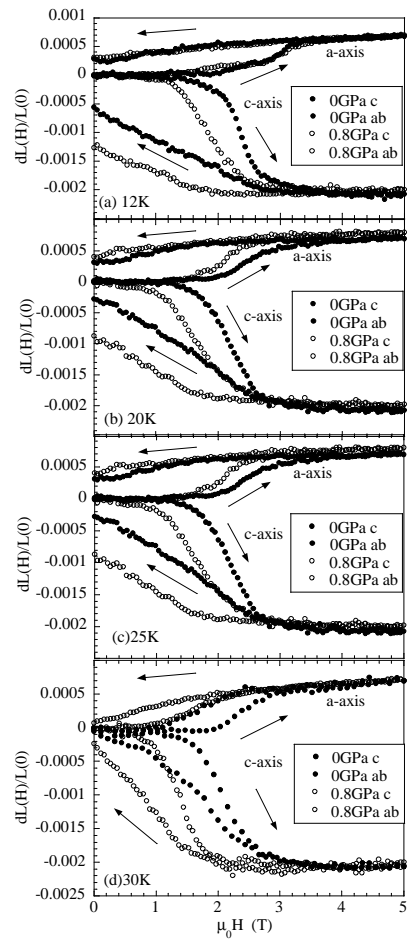}%
\caption{The $ab$-plane and $c$-axis magnetostrictions,  $dL_{a}(H)/L_{a}(0)$ and  $dL_{c}(H)/L_{c}(0)$, both under ambient pressure and a hydrostatic pressure of 0.8 GPa, at different temperatures (a)12K,(b)20K,(c)25K and (d) 30K. 
The applied field  is parallel to the $c$-axis ($H\parallel c$). }
\label{}
\end{figure}%
Figure 2 shows the $ab$-plane and $c$-axis magnetostrictions,  $dL_{a}(H)/L_{a}(0)$ and  $dL_{c}(H)/L_{c}(0)$, both under ambient pressure and a hydrostatic pressure of 0.8 GPa at different temperatures, where the  applied field is parallel to the $c$-axis (H//$c$). Here, the value of $dL_{i}(H)$ is defined as $L_{i}(H)-L_{i}(0)$. 
First, we observe that the $c$-axis rapidly shrinks near 2T upon applying the field, but the  $a$-axis expands at the same field value.
The field-induced IM transition accompanies  a stable decrease of the $c$-axis  by $\sim 0.2\%$, in contrast with a small increase of the $a$-axis  by $\sim 0.06\%$, resulting in a volume shrinkage of $\sim -0.08\%$ 
($dV(H)/V=2dL_{a}(H)/L_{a}(0)+dL_{c}(H)/L_{c}(0)$ \cite{GA00}). This value is in good agreement with the volume striction
  $dV/V\sim -0.09\%$ associated with the spontaneous IM transition of the parent compound La$_{1.2}$Sr$_{1.8}$Mn$_{2}$O$_{7}$,indicating that a volume shrinkage in the metallic state is  a consequence of the charge delocalization
\cite{IB95,ME99}. Here, the variations of the lattice parameters of the parent manganite,$\Delta a/a\sim -0.08\%$ and $\Delta c/c\sim 0.07\%$, are
estimated from neutron diffraction measurements \cite{AR97}. 
At selected temperatures, a clear hysteresis in the magnetostriction curves was observed, which has
a close relationship to a memory effect in magnetoresistance,magnetization and magnetothermal conduction of the $z=0.6$crystal \cite{AP01,GO01,MA03}.
Next, the application of pressure to the magnetostrictions  reduces the critical fields  and also enhances the remanent magnetostriction
just after removing the applied field.
We show in Fig.3 the pressure effect on the $ab$-plane and $c$-axis magnetostrictions,  $dL_{a}(H)/L_{a}(0)$ and  $dL_{c}(H)/L_{c}(0)$
in the case of the field applied in the $ab$ plane($H\parallel ab$). 
The magnetostriction behavior in $H\parallel ab$ is qualitatively similar to that in $H\parallel c$.
We note that quantitative differences in $dL/L$ between $H\parallel ab$ and $H\parallel c$ are higher critical fields and larger hysteresis regions in the former case. This finding is probably related to the easy axis of magnetization 
through the orbital occupation of $e_{g}$ electron between the $d_{x^2-y^2}$ and  $d_{3z^2-r^2}$ states as mentioned below. 
\begin{figure}[ht]
\includegraphics[width=8cm]{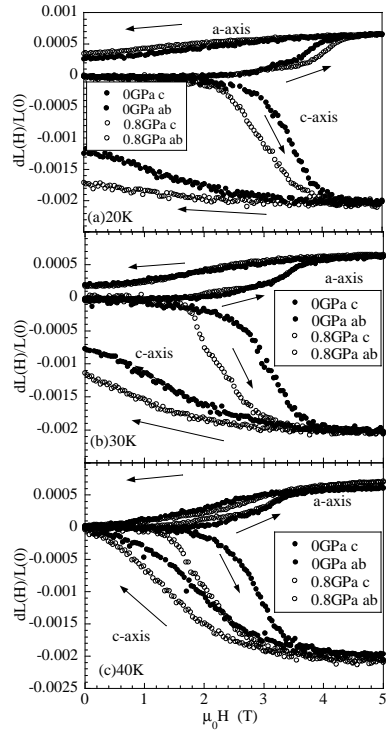}%
\caption{The $ab$-plane and $c$-axis magnetostrictions,  $dL_{a}(H)/L_{a}(0)$ and  $dL_{c}(H)/L_{c}(0)$, both under ambient pressure and a hydrostatic pressure of 0.8 GPa, at different temperatures (a)20K,(b)30K,and (c)40K. 
The applied field is parallel to the $ab$-plane ($H\parallel ab$). }
\end{figure}%

\begin{figure}[ht]
\includegraphics[width=9cm]{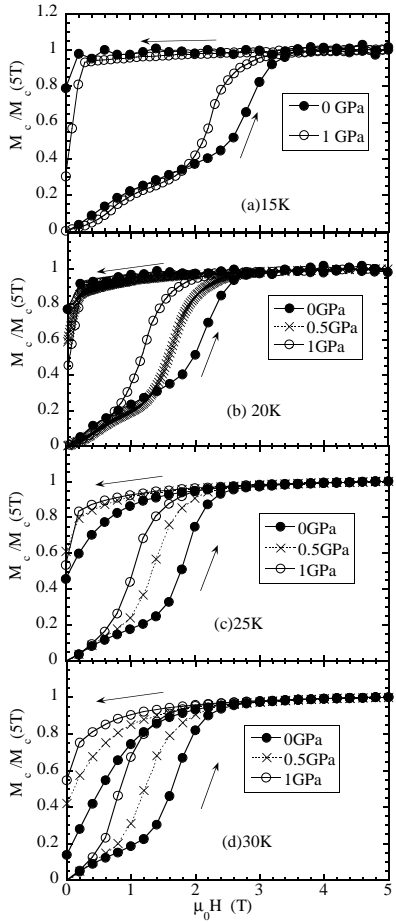}%
\caption{$c$-axis magnetization under different pressures (0,0.5, and 1.0 GPa)at selected temperatures (a)15K,(b)20K,(c)25K and (d) 30K. The data are normalized by the saturated magnetization at 5T.  }
\label{mc}
\end{figure}%

\begin{figure}[ht]
\includegraphics[width=9cm]{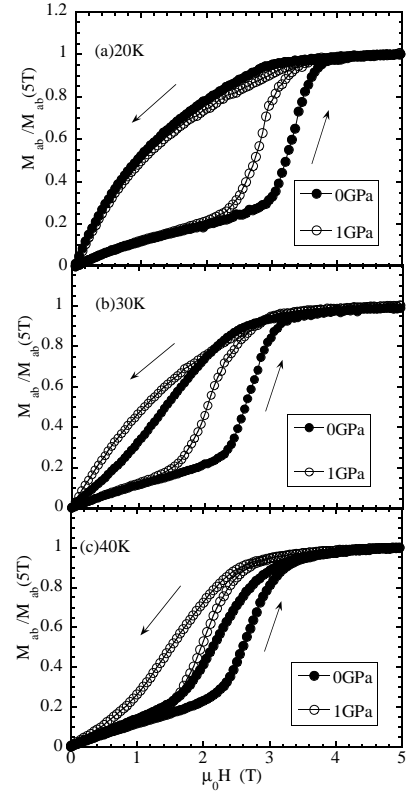}%
\caption{$ab$-plane magnetization under different pressures (0 and 1.0 GPa)at selected temperatures (a)20K,(b)30K,and (c) 40K. The data are normalized by the value of $M_{ab}$ at 5T.   }
\label{mab}
\end{figure}%
Now, we examine the $c$-axis magnetization under different pressures (0,0.5, and 1.0 GPa) as shown in Fig.\ref{mc}.
For comparison, the $ab$-plane magnetization data are given in Fig.\ref{mab}.
Here, the value of $M_{c}$(5T) reaches $\sim$ 3.5$\mu_{B}$/Mn site, close to the full moment of 3.6$\mu_{B}$/Mn site
while the value of  $M_{ab}$ is $\sim$ 3.0$\mu_{B}$/Mn site, which is by about 20 $\%$ smaller than the ideal value \cite{ AP01}.  The difference in the saturated magnetizations  between $M_{ab}$ and $M_{c}$ at 5T probably arises from the easy axis of magnetization along the 
$c$-axis associated with the rise of the  $d_{3z^2-r^2}$  orbital occupancy at $z$=0.6 \cite{WA03}.
This finding is naturally understood by considering the importance of the spin-orbit interaction in the bilayered manganite system \cite{MA00}.
We note that  the $ab$-plane magnetization exhibits no remanent value at all temperatures,
 in striking contrast with the substantial values of remanent magnetostriction,remanent magnetoresistance
and remanent magnetothermal conduction\cite{AP01,MA03}. 
This discrepancy is not the central issue discussed here, but it has a close relationship with the formation
of magnetic domains,keeping their local moments \cite{TO05}. 
First, the application of pressure on magnetization  suppresses  the critical field inducing the PMI to FMM transition to the system, as well as the pressure effect on magnetostriction.
The phase transition lines in Fig.1 are defined as a maximum of $dM/dH$ because the $MH$ curves show  metamagnetic
behavior.  At higher temperatures (30K), it is true that the remanent magnetization is also enhanced upon increasing the pressure. However, at lower temperatures (15K and 20K)  the remanent $M_{c}$ shows a rapid decrease with the applied pressures.
The negative pressure effect on remanent $M_{c}$, at lower $T$, contrasts with the positive pressure dependence of remanent magnetostriction at the corresponding $T$.  
Recently, neutron diffraction studies under pressure on bilayered manganite crystal with the same composition 
have shown similar results for the pressure influence on magnetization \cite{GU05}.
We  comment now on the pressure effect on the MnO$_{6}$ octahedron sites in the parent bilayered manganite La$_{1.2}$Sr$_{1.8}$Mn$_{2}$O$_{7}$. Argyriou et al. have reported that upon increasing pressure up to 0.6GPa,
in the PMI state, both the Mn-O(1) and Mn-O(3) bond lengths shrink but the Mn-O(2) bond length elongates \cite{AR97L}. Here, the O(1) and O(2) oxygen atoms are located at the apical site along the $c$-axis,where O(1) lies between two MnO$_{2}$ layers and O(2) within a rocksalt-type La/Sr-O layer.
The O(3) oxygen atom is within the MnO$_{2}$ layer . 
Assuming such a variation of the MnO$_{6}$ octahedron in the PMI state of Pr-substituted La$_{1.2}$Sr$_{1.8}$Mn$_{2}$O$_{7}$, the Mn-O(1)-Mn interactions along the $c$-axis and the Mn-O(3)-Mn interactions in the  $ab$-plane are expected
to be strengthened upon application of pressure. In other words, the double exchange (DE) interaction between Mn$^{3+}$ and Mn$^{4+}$
within the MnO$_{2}$ single layer and the DE interaction along the $c$-axis within the bilayer are enhanced and this is closely
related to the decrease of the critical fields of the FMM state with pressure. 
On the other hand, the reciprocal response of Mn-O(2) to the applied pressures causes an elongation of the distance between adjacent bilayers,
and it thus weakens the super-exchange interactions between adjacent bilayers, keeping the DE interaction in the bilayer.
This assumption  describes well the negative pressure effect on the remanent $M_{c}$ at lower temperatures
and is consistent with the positive effect on the remanent magnetostriction.
Accordingly, the DE interaction-driven FMM state within a bilayer is enhanced with pressure but the SE interaction-driven ferromagnetic state between adjacent bilayers is suppressed \cite{GU05}.
\begin{figure}[ht]
\includegraphics[width=9cm]{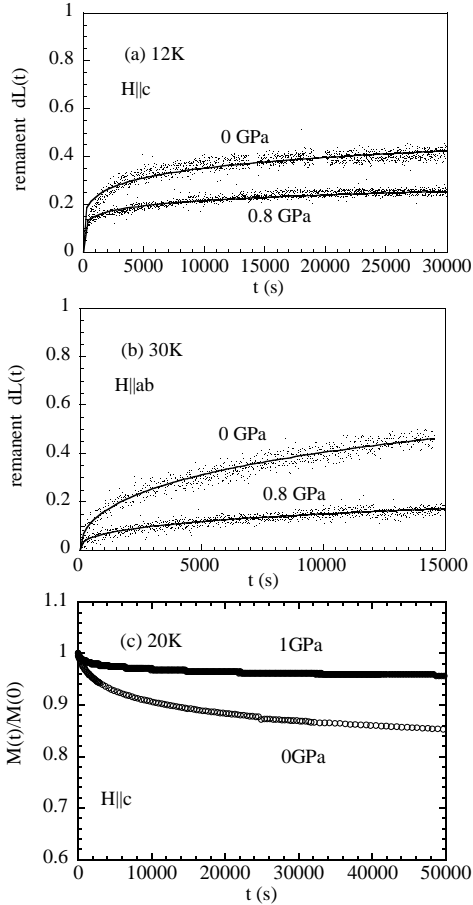}%
\caption{(a)Normalized remanent magnetostriction, 
$d(c/a)/(c/a)$
as a function of time at 12K, just after the field along the $c$-axis was switched off ($H\parallel c$). The solid lines correspond to  a fit to the data points by  a stretched exponential function 1-exp$[-(t/\tau)^\beta]$, 
where  $\tau$ and $\beta$ represent the  characteristic 
 relaxation  time  and exponent, respectively. 
In this case , we get $\tau$=$4.0\times10^5$ s with $\beta$=0.22 and
$\tau$=$3.7\times10^7$ s with $\beta$=0.17, at 0 and 0.8 GPa, respectively.
(b) The  normalized remanent magnetostriction at 30K, just after the field in the $ab$-plane was switched off ($H\parallel ab$).
$\tau$=$4.0\times10^4$ s with $\beta$=0.47 at 0 GPa and $\tau$=$1.4\times10^6$ s with $\beta$=0.37 at 0.8 GPa.
(c)The  normalized remanent magnetization at 20K, just after the field along the $c$-axis was switched off ($H\parallel c$)
$\tau$=$3.1\times10^6$ s with $\beta$=0.16 at 0 GPa and $\tau$=$3.4\times10^9$ s with $\beta$=0.1 at 1.0 GPa.
The data points are fitted using $M(t)/M(0)=$1-exp$[-(\tau /t)^\beta ]$.
}
\label{rema}
\end{figure}%

\begin{figure}[ht]
\includegraphics[width=9cm]{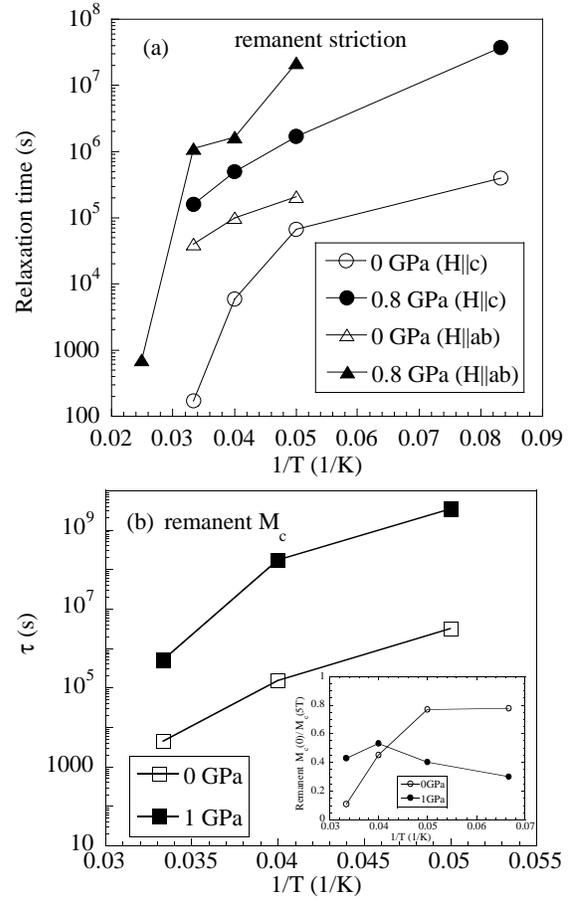}%
\caption{(a)Pressure effect on the relaxation time of the remanent lattice striction as a function of $1/T$. (b) 
Pressure effect on the relaxation time of  
the remanent $c$-axis magnetization.  For comparison, the inset represents the remanent $M_{c}$ versus $1/T$
taken from the data in Fig.\ref{mc}. }
\label{relax}
\end{figure}%

Finally, we examine the  pressure effect on the relaxation time of  remanent magnetostriction.
Just after the removal of the applied fields, both the $ab$-plane and the $c$-axis remanent magnetostrictions
are recorded as a function of time. 
The  normalized remanent magnetostriction, $d(c/a)/(c/a)$, is shown as a function of time in Fig.\ref{rema}
where  the anisotropic lattice striction $d(c/a)/(c/a)$ is estimated from the $ab$-plane and $c$-axis remanent data
using $dL_{c}/L_{c}-dL_{a}/L_{a}$.
The data points are normalized  in such a manner that we make the initial value, just after the removal of field, to be zero, while the virgin value before the application of field becomes unity.
In our case, the lattice variation along the $c$-axis is by a factor of $\sim$ 3  as large as the value in the $ab$-plane and
the value of $d(c/a)/(c/a)$ is almost taken as $dL_{c}/L_{c}$.
As reported in previous papers\cite{GO01,MA04,MA05}, 
the temporal profile of the remanent magnetization and magnetostriction follows a stretched exponential function 
1-exp$[-(t/\tau)^\beta]$, where  $\tau$ and $\beta$ represent the  characteristic 
relaxation  time  and exponent, respectively. 
The stretched exponential behavior of the relaxation curves indicates the existence of  frustrations among competing interactions such as the FM and AFM interactions in spin glass systems.
We believe that the remanent magnetostriction shows a stretched exponential decay  because of  the competition between the double exchange 
and JT type lattice-orbital interactions\cite{MA04}.
The former interaction causes a suppression of the local lattice distortion along the $c$-axis through the itinerant state, 
but the latter favors  lattice deformation through the local Jahn-Teller effect\cite{ME99}.
Neutron scattering measurements on the parent bilayered manganite La$_{2-2x}$Sr$_{1+2x}$Mn$_{2}$O$_{7}$($x$=0.38) have revealed 
the presence of an orbital frustration in the paramagnetic insulating state which prevents the system from the formation
of a long-range orbital- and charge-ordered state \cite{AR02}. We expect that this type of  frustration survives for the Pr-substituted crystal with nearly same doping and is relevant to the  phenomenon of slow  relaxation observed in remanent striction. 

For a stretched exponential fit to the data points at 12K, we get $\tau$=$4.0\times10^5$ s under ambient pressure.
Under a pressure of 0.8GPa,  $\tau$ becomes $3.7\times10^7$ s which is much longer than the relaxation time without pressure.
As a result, in Fig.\ref{relax} , we summarize the relaxation time of both the remanent striction and remanent magnetization as a function of $1/T$.
The application of pressures up to $\sim$ 0.8GPa to the system increases the relaxation time of the lattice by about two orders of magnitude, giving a more stable metallic state coupled with a suppression of MnO bond lengths in  the MnO$_{6}$ local lattice. 
The relaxation data  of remanent magnetization under a pressure of 1GPa at 20K  are presented in Fig.\ref{rema}(c). 
In spite of the negative pressure effect on the remanent $M_{c}$ at lower $T$, as mentioned before( Fig.\ref{mc} (a) and (b)),
the relaxation time is strongly enhanced at 1GPa by more than two orders of magnitude and shows a positive
pressure dependence.
The difference in the relaxation time between remanent magnetization and magnetostriction 
can probably be explained by the fact that the magnetostriction observed here is not  associated with a long-range order 
parameter such as the long-range orbital order reported in the $x$=0.5 crystal.
On the other hand, the magnetization is closely related to it \cite{MA05}.

\section{SUMMARY}
 We have demonstrated  the effect of pressure on magnetostriction, both in the $ab$-plane and along the $c$-axis , in a bilayered manganite single crystal of (La,Pr)$_{1.2}$Sr$_{1.8}$Mn$_{2}$O$_{7}$ over the temperature region where  the field-induced ferromagnetic metal (FMM) transition takes place . 
For comparison, we have examined the pressure dependence of  the magnetization curves  at the corresponding temperatures.
The applied pressure reduces the critical field of the FMM transition  and it  also enhances the remanent magnetostriction. 
The quantitative differences observed in $dL/L$ between $H\parallel ab$ and $H\parallel c$ are higher critical fields and larger hysteresis regions in the former case, which is probably related to the easy axis of magnetization along the $c$-axis through the orbital occupation of $e_{g}$ electron between the $d_{x^2-y^2}$ and  $d_{3z^2-r^2}$ states. 
An anomalous pressure effect on the remanent lattice relaxation is observed to be  similar to the pressure effect on the remanent magnetization along the $c$-axis. 
In spite of the negative pressure effect on remanent $M_{c}$ at lower $T$,
the relaxation time is strongly enhanced at 1GPa by more than two orders of magnitude and shows a positive
pressure dependence.
These findings are understood considering that 
the  double-exchange interaction-driven  FMM state within a bilayer is strengthened upon application of pressure.
Application of pressure reinforces the DE interaction, both in the $ab$-plane and along the $c$-axis, through
a shrinkage of Mn-O(1) and Mn-O(3) bond lengths, resulting in a giant effect on the relaxation time of remanent
striction and remanent magnetization.
We believe that the slow relaxation observed here correlates to the orbital frustration existing 
in the paramagnetic insulating state of parent bilayered manganite.

\begin{acknowledgments}

 This work was partially supported by a Grant-in-Aid for Scientific Research from Japan Society of the Promotion
of Science. We thank H.Noto for his technical support. One of the authors (M.Matsukawa) would like to thank T.Naka and Dr.A.Matsushita, NIMS for their help in pressure measurement on magnetostriction. 
\end{acknowledgments}


\begin{thebibliography}{30}

\bibitem{TO00} \textit{Colossal Magnetoresistive Oxides}, edited by Y.Tokura (Gordon and Breach,New York,2000).
\bibitem {ZE51}C.Zener,Phys.Rev.82,403(1951); P.G.deGennes,$ibid$.118,141 (1960).
\bibitem {MI95} A.J.Millis,P.B.Littlewood, and B.I.Shraiman,
Phys.Rev.Lett.74,5144(1995); A.J.Millis, B.I.Shraiman, and R.Mueller, $ibid$.77,175 (1996).
\bibitem {DA01} For a recent review, see E.Dagotto, T.Hotta, and A.Moreo, 
Phys.Rep.344,1 (2001).
\bibitem {MO96}Y.Moritomo, A.Asamitsu, H.Kuwahara, and Y.Tokura,
 Nature 380,141 (1996).
\bibitem {MO97}Y.Moritomo, Y.Maruyama,T.Akimoto, and A.Nakamura
, Phys.Rev.B56,R7057(1997).
\bibitem{OG00}H.Ogasawara,M.Matsukawa,S.Hatakeyama,M.Yoshizawa,
M.Apostu, R.Suryanarayanan, G.Dhalenne, A.Revcolevschi, K.Itoh, and N.Kobayashi, 
J.Phys.Soc.Jpn.69,1274(2000).
\bibitem{HW95}H.Y.Hwang,S-W.Cheong,P.G.Radaelli,M.Marezio,and B.Batlogg,
Phys.Rev.Lett.75,914(1995).
\bibitem{AR97L}D.N.Argyriou,J.F.Mitchell,J.B.Goodenough,
O.Chmaissem,S.Short,and J.D.Jorgensen,
Phys.Rev.Lett.78,1568(1997).
\bibitem{KI97}T.Kimura,A.Asamitsu,Y.Tomioka,and Y.Tokura,
Phys.Rev.Lett.79,3720(1997).
\bibitem{ZH98}J.S.Zhou,J.B.Goodenough,and J.F.Mitchell, 
Phys.Rev.B58,R579(1998).
\bibitem{AR00}D.N.Argyriou,H.N.Bordallo,B.J.Campbell,A.K.Cheetham,
D.E.Cox,J.S.Gardner,K.Hanif
,A.dos Santos,and G.F.Strouse,Phys.Rev.B61,15269(2000).
\bibitem{AR02}D.N.Argyriou,J.W.Lynn,R.Osborn,B.Campbell,J.F.Mitchell,
U.Ruett,H.N.Bordallo,A.Wildes,and C.D.Ling,
Phys.Rev.Lett.89,036401(2002).
\bibitem {AP01}M.Apostu, R.Suryanarayanan, A.Revcolevschi, 
H.Ogasawara, M.Matsukawa, M.Yoshizawa, 
and N.Kobayashi,Phys.Rev.B64,012407(2001).

\bibitem{GA00} B.Garcia-Landa,C.Marquina,M.R.Ibarra,
G.Balakrishnan,M.R.Lees,
and D.McK.Paul,
Phys.Rev.Lett.84,995(2000).
\bibitem{IB95}M.R.Ibarra,P.A.Algarabel,C.Marquina,J.Blasco, and J.Garcia
Phys.Rev.Lett.75,3541(1995).
\bibitem {ME99}M.Medarde,J.F.Mitchell,J.E.Millburn, S.Short, 
and J.D.Jorgensen, Phys.Rev.Lett.83,1223(1999).
\bibitem{AR97}D.N.Argyriou,J.F.Mitchell,C.D.Potter,S.D.Bader,R.Kleb,and J.D.Jorgensen 
,Phys.Rev.B55,R11965(1997).

\bibitem {GO01}I.Gordon,P.Wagner,V.V.Moshchalkov,Y.Bruynseraede,
M.Apostu, R.Suryanarayanan,and A.Revcolevschi,
Phys.Rev.B64,092408(2001).
\bibitem {MA03}M.Matsukawa,M.Narita,T. Nishimura,M.Yoshizawa, M.Apostu,R.Suryanarayanan,
A.Revcolevschi,K.Itoh, and N.Kobayashi,, Phys.Rev.B67,104433(2003).
\bibitem {WA03}F.Wang,A.Gukasov,F.Moussa,M.Hennion, M.Apostu,
R.Suryanarayanan, and A.Revcolevschi, Phys.Rev.Lett.91,047204(2003).
\bibitem{MA00}R.Maezono, and N.Nagaosa,Phys.Rev.B61,1825(2000). 
\bibitem {TO05}Y.Tokunaga,M.Tokunaga, and T.Tamegai, 
Phys.Rev.B71,012408(2005).
\bibitem{GU05}A.Gukasov,F.Wang,B.Anighoefer,Lunhua He,R.Suryanarayanan,and A.Revcolevschi,
Phys.Rev.B72,092402(2005).

%
\bibitem {MA04}M.Matsukawa,M.Chiba,K.Akasaka, R.Suryanarayanan, 
 M.Apostu, A.Revcolevschi,S.Nimori, and N.Kobayashi, 
Phys.Rev.B70,132402(2004).
\bibitem {MA05}M.Matsukawa,K.Akasaka,H.Noto, R.Suryanarayanan, 
S.Nimori,M.Apostu, A.Revcolevschi, and N.Kobayashi, 
Phys.Rev.B72,064412(2005).



 
\end{thebibliography}
\end{document}